\documentclass[draft]{agujournal2019}
\usepackage{url} 
\usepackage{lineno}
\usepackage{amsmath}
\newcommand{\Mod}[1]{\ (\mathrm{mod}\ #1)}
\draftfalse

\journalname{Geophysical Research Letters}

\begin{document}

%
%


\title{The application of Convolutional Neural Networks to Detect Slow, Sustained Deformation in InSAR Timeseries}

%
%




\authors{N. Anantrasirichai\affil{1}, J. Biggs\affil{2}, F. Albino\affil{2}, and D. Bull\affil{1}}

\affiliation{1}{Visual Information Laboratory, University of Bristol, UK}
\affiliation{2}{School of Earth Sciences, University of Bristol, UK}




\correspondingauthor{N. Anantrasirichai}{n.anantrasirichai@bristol.ac.uk}




\begin{keypoints}
\item We adapt a pre-trained convolutional neural networks for detecting slow, steady deformation in InSAR images.
\item Synthetic tests show the detection threshold is ~6.3 cm in volcanic environments, but is reduced to 5.0 cm by decreasing wrapping interval.
\item We demonstrate the method on cumulative timeseries of interferograms from Campi Flegrei, Italy (8.5 cm/yr) and Dallol, Ethopia (3.5 cm/yr).
\end{keypoints}

%
%


\begin{abstract}

Automated systems for detecting deformation in satellite InSAR imagery could be used to develop a global monitoring system for volcanic and urban environments. Here we explore the limits of a CNN for detecting slow, sustained deformations in wrapped interferograms. Using synthetic data, we estimate a detection threshold of 3.9cm for deformation signals alone, and 6.3cm when atmospheric artefacts are considered. Over-wrapping reduces this to 1.8cm and 5.0cm respectively as more fringes are generated without altering SNR. We test the approach on timeseries of cumulative deformation from Campi Flegrei and Dallol, where over-wrapping improves classication performance by up to 15\%. We propose a mean-filtering method for combining results of different wrap parameters to flag deformation. At Campi Flegrei, deformation of 8.5cm/yr was detected after 60days and at Dallol, deformation of 3.5cm/yr was detected after 310days. This corresponds to cumulative displacements of 3cm and 4cm consistent with estimates based on synthetic data.

\end{abstract}

%
%

\section{Introduction}
Interferometric synthetic aperture radar (InSAR) can be used to measure ground displacement over large geographic areas. However, while detecting deformation using InSAR images is conceptually straightforward, it is difficult to automate. For volcanic regions, atmospheric water vapour is a particular challenge since the stratification of atmospheric water vapour forms a topographically correlated pattern which can be difficult to distinguish from the deformation caused by the pressurisation of magma reservoirs beneath volcanic edifices \cite{Beauducel:volcano:2000, Ebmeier:Applicability:2013, Remy:revised:2015}. In urban environments, the deformation patterns typically have at a lower rate and smaller spatial scale, but are more distinct from atmospheric noise. 

Previous studies have shown that deep convolutional neural networks (CNNs) have the capability to identify volcanic deformation signals from a large dataset of wrapped interferograms \cite{Anantrasirichai:Application:2018,Anantrasirichai:deep:2019,Valade:towards:2019}. The high-frequency content of wrapped fringes provides strong features for machine learning algorithms and \citeA{Anantrasirichai:Application:2018,Anantrasirichai:deep:2019} use a method based on edge detection.  The outputs are expressed as a probability, which can be used to flag deformation. However, for C-band satellites, one fringe corresponds to 2.8 cm of deformation, or 1.80~m/yr for a 12-day interferogram and 3.60~m/yr for 6-day interferogram. The approach of \citeA{Valade:towards:2019} also only tested short-term interferograms that showed deformation of $>$10 cm \cite{GVP2019}. These high rates are typically only observed for very short periods associated with dyke intrusions or eruptions \cite{biggs2017global}.  Yet there are many deformation signals that occur at lower rates, but for longer duration, such as sustained uplift at silicic volcanoes  \cite{Trasatti:uplift:2008,Remy:persistent:2014,Montgomery:renewed:2015,Henderson:time:2017,Lloyd:Evidence:2018}, subsidence and heave in former coalfields \cite{McCay:meta:2018}, engineering projects such as tunnelling and landsliding of natural and engineered slopes \cite{Whiteley:Geophysical:2019}. 

Here we investigate two adaptations to enable a machine learning system to detect slow, sustained deformation: 1) the use of a daisy-chain of interferograms to increase the time interval and hence the signal to noise ratio (SNR) and 2) rewrapping the interferograms to generate additional fringes, but without altering the SNR.  

\section{Methods}


\subsection{Convolutional Neural Networks}

Convolutional neural networks (CNNs) are deep feed-forward artificial neural networks, designed to take advantage of 2D structures, such as an image. In this study, we use a transfer-learning strategy, employing the fine-tuned AlexNet developed by \citeA{Anantrasirichai:deep:2019}.  AlexNet contains five convolutional layers and three fully connected layers. The first, the second and the fifth convolutional layers are followed by max-pooling layers. ReLU is applied after every convolutional and fully connected layer.  The interferogram is first converted into a grayscale image (i.e. the pixel values are scaled to $[0,255]$) and divided into overlapping patches at the required input size for AlexNet (224$\times$224 pixels).  Following \citeA{Anantrasirichai:deep:2019}, the top-left position of each patch is then repeatedly shifted by 28 pixels to cover the entire image. The output of the prediction process is a probability $P$ of there being deformation in each patch and the probabilities from overlapping patches are merged using a rotationally symmetric Gaussian lowpass filter with a size of 20 pixels and standard deviation of 5 pixels. The network was initially trained with synthetic InSAR data representing deformation, turbulent and stratified atmospheric contributions and then retrained with a combined dataset consisting of synthetic models and selected real examples. The interferogram is flagged as containing deformation when $P>0.5$.

\subsection{InSAR Dataset}

Initially we test the detection threshold of the network using synthetic data. We create a dataset of 405 synthetic interferograms ($X$) using 3 components, namely deformation $D$, stratified atmosphere $S$, and turbulent atmosphere $T$, using the linear function $X = D+\alpha S+\beta T$, where $\alpha, \beta \in \{0, 0.5, 1\}$.
Synthetic deformation signals, $D$, are generated using a point pressure source (Mogi) model \cite{Mogi:Relation:1958}, which demonstrates surface deformation associated with inflation and deflation of a magma chamber. In this study, $D$ is modelled with depths of 3, 4 and 5 km, incidence angles of 1$^\circ$, 23$^\circ$ and 44$^\circ$, and volumes of 10$^\nu$~m$^3$, where $\nu \in [5, 5.5, 6, ..., 7]$.
For stratified atmosphere $S$, we use the signal at Beru volcano (20170104-20170317) via
the Generic Atmospheric Correction Online Service (GACOS). It is based on weather model data \cite{yu2018generic} and a zenith total delay (ZTD) map is derived from the high-resolution water vapour delays generated by the European Centre for Medium-Range Weather Forecasts (ECMWF).
Turbulent atmospheric delays, $T$, are spatially correlated and their covariance can be described using an exponentially decaying  function.  The one-dimensional covariance function is $c_{ij} = \sigma^2_{max} e^{(-\kappa d_{ij})}$, where $c_{ij}$ is the covariance between pixels $i$ and $j$, $d_{ij}$ is the distance between the pixels, $\sigma^2_{max}$ is maximum covariance and $\kappa$ is the decay constant, which is equivalent to the inverse of the $e$-folding wavelength \cite{biggs2007multi}. We employ $\sigma^2_{max}$ of 7.5~mm$^2$ and $\kappa$ of 8~km \cite{Anantrasirichai:deep:2019}.

We then test our approach on two examples: Campi Flegrei (Italy) where a new pulse of uplift began in July 2017, and Dallol (Ethiopia), which is uplifting at a steady rate of 3.5 cm/yr (Figure \ref{fig:result_real_plot}).  
Campi Flegrei is an active 13-km wide caldera located within the bay of Naples, a dense populated area of 3 millions inhabitants \cite{global2013volcanoes}. Since the last eruption in 1538 at Monte Nuovo, the caldera system showed signs of unrest with episodes of rapid ground uplift (e.g. 1950-52, 1969-72 and 1982-84) followed by long period of recovery associated with slow ground subsidence (e.g. 1984-2010) \cite{delgaudio2010unrest, troiano2011ground}. Dallol is an Ethiopian volcano located in the Danakil depression. The crater lies 48 meters below the sea level and was formed during a phreatic eruption in 1926 \cite{global2013volcanoes}. Recent unrest associated with the emplacement of a magma intrusion have been detected by InSAR in 2004 \cite{nobile2012dike}, and a phreatic eruption have been recorded on January 2011 \cite{global2013volcanoes}.

For each volcano, we process Sentinel-1 interferograms using the LiCSAR processor (http://comet.nerc.ac.uk/COMET-LiCS-portal/) and perform a least-squared inversion on the unwrapped interferograms dataset to obtain the time series of ground deformation \cite{schmidt2003time, usai2003least}.

\subsection{Data Wrapping}

The phase $\psi$ of an interferogram varies between $-\pi$ and $\pi$, and is typically unwrapped before geophysical analysis \cite{Chen:phase:2002}. CNNs have been previously applied to wrapped InSAR data \cite{Anantrasirichai:Application:2018, Valade:towards:2019}, which is preferable for detecting rapid deformation signals as the unwrapping process is time-consuming and can introduce errors. However, for studying slow, sustained deformation, unwrapping is necessary to produce time-series of cumulative deformation. Nonetheless, fringes provide strong features for machine learning and in this study, we chose to re-wrap the cumulative time series and use the pre-trained network of \citeA{Anantrasirichai:deep:2019}. 

We test whether altering the number and location of the wrap boundaries has the potential to improve the ability of the CNN to detect deformation. Reducing the wrapping interval increases the number of fringes without altering the signal to noise ratio. We generate a new phase $\psi'_\mu$ with a wrap gain $\mu$ using $\psi'_\mu \equiv {\mu}  \psi  \Mod{2\pi}$, where $\mu$ is a positive integer. The wrap interval is reduced by $1/\mu$ of the original phase value, and the number of fringes increases $\mu$ times, e.g. $\mu$=2 produces twice as many fringes (Figure \ref{fig:result_syn_mul_shift_subj} left). We also consider shifting the wrapping boundaries by adding a constant phase offset $\tau$ to the unwrapped interferogram prior to rewrapping using $\psi'_\tau$, i.e. $\psi'_\tau  \equiv   \psi + \tau  \Mod{2\pi}$. The phase discontinuities occur in physically arbitatrary locations, so for some cases, the number of fringes will increase, but in others it will decrease (Figure \ref{fig:result_syn_mul_shift_subj} right). 

\section{Results}
\subsection{Detection Thresholds}
\begin{figure*}[!ht]
	\centering
      		\includegraphics[width=\textwidth]{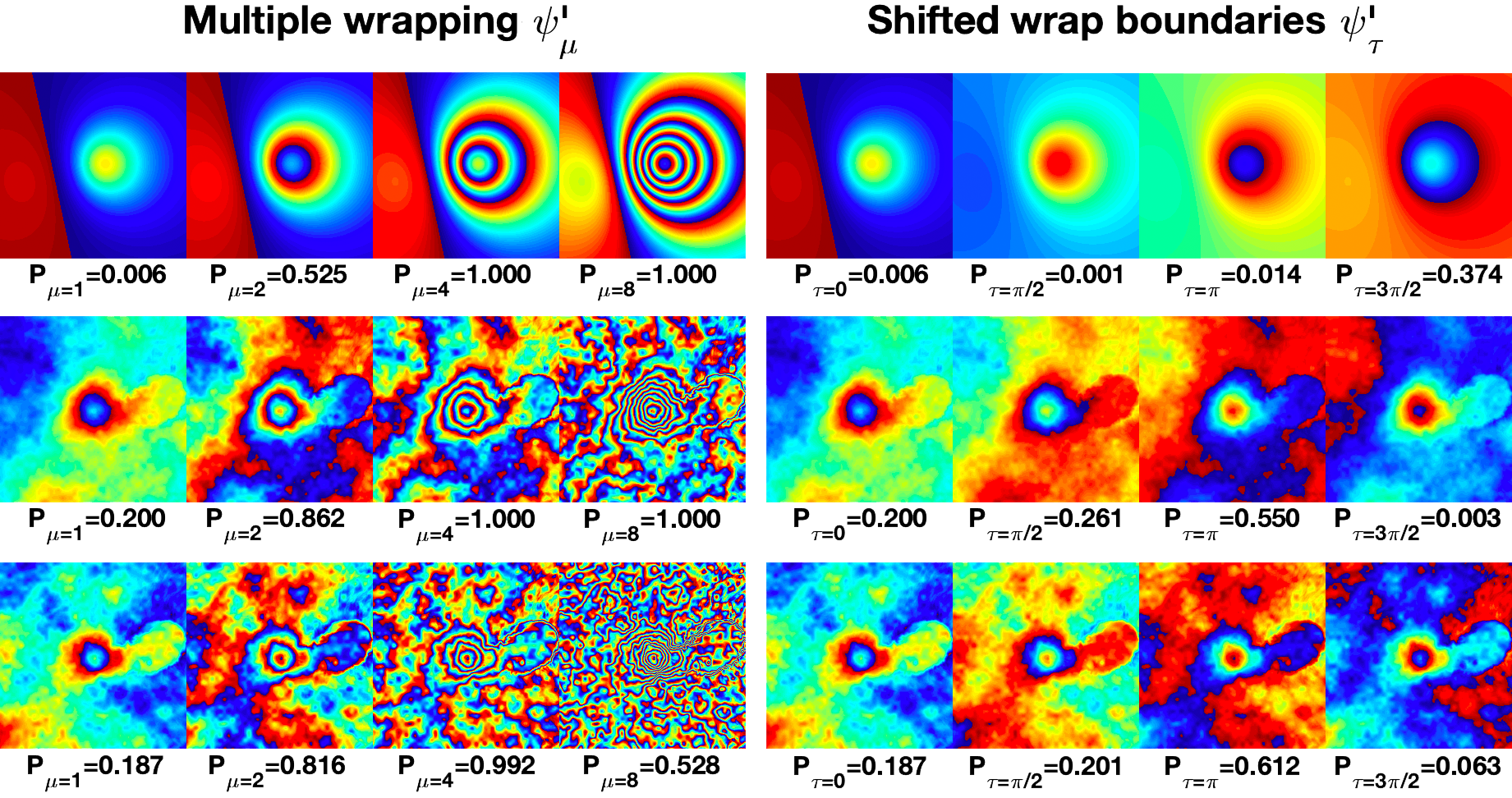}  
					\caption{Example synthetic interferograms re-wrapped with $\mu$=1, 2, 4 and 8 (left column)  and with shifted wrap boundaries with $\tau$= 0, $\pi$/2, $\pi$ and 3$\pi$/2 (right column). The relative weighting of the stratified and turbulent components are given by $\alpha$ and $\beta$ respectively. Top row $\alpha$=$\beta$=0, middle row $\alpha$=$\beta$=0.5; bottom row $\alpha$=$\beta$=1. The probability, $P$, that the image contains a deformation signal according to the CNN is listed beneath each image.}
    \label{fig:result_syn_mul_shift_subj}
\end{figure*}

Figure \ref{fig:result_syn_mul_shift_subj} left shows examples of synthetic interferograms with wrap gains $\mu$=1, 2, 4 and 8. For each example, the CNN output probability is below the detection threshold ($P<0.5$) at $\mu=1$, but greater than the threshold ($P>0.5$) at $\mu$=2, 4 and 8. This illustrates that the increased number of fringes improves the ability of the CNN to detect the deformation, even though the signal to noise ratio of the inteferograms remains constant. In contrast, shifting wrap boundaries does not always improve the detection performance as shown in Figure \ref{fig:result_syn_mul_shift_subj} right, where $P$ may increase and also decrease.

\begin{figure*}[!ht]
	\centering
      		\includegraphics[width=\textwidth]{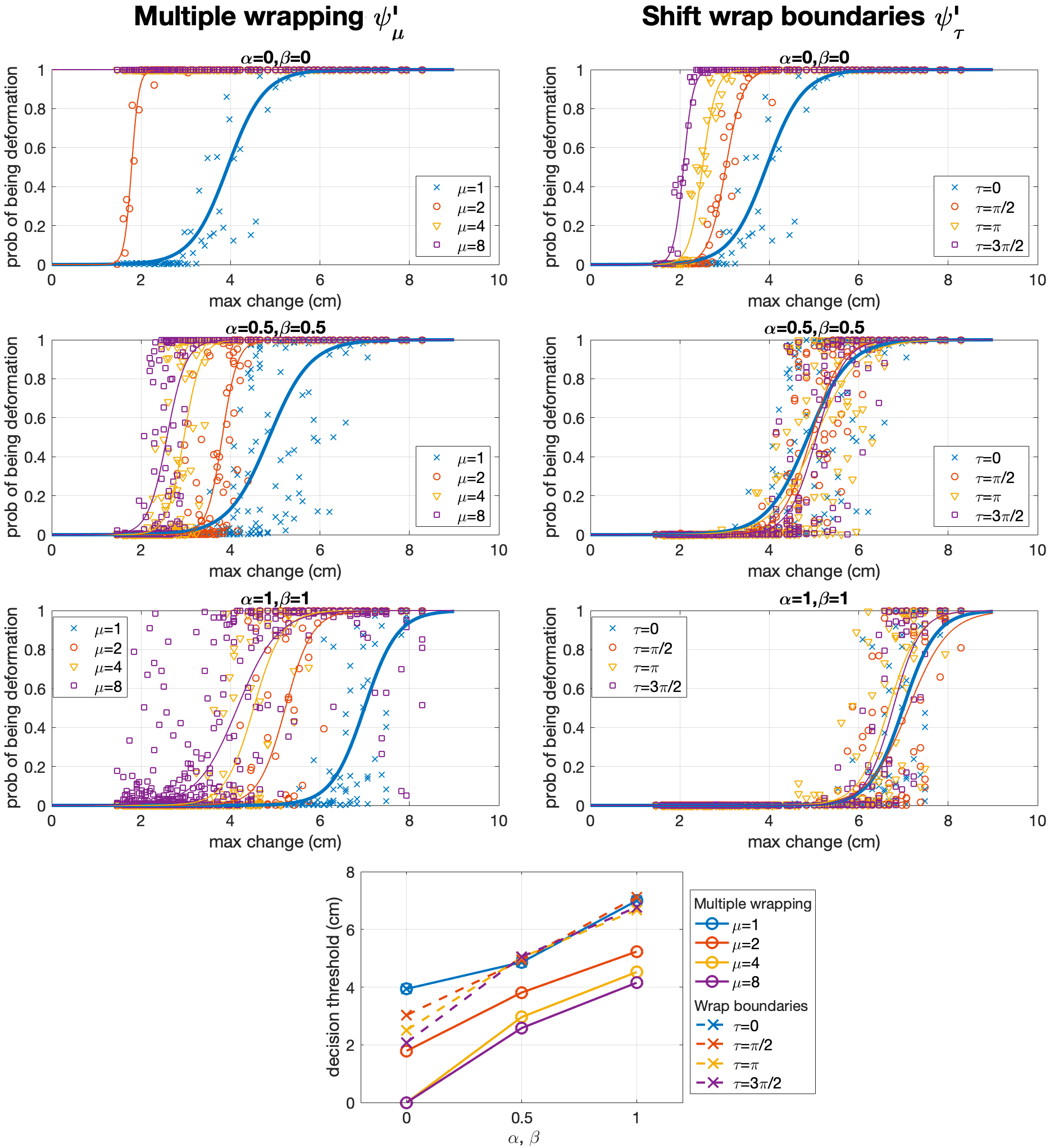}  
					\caption{Analysis of the detection threshold of the CNN using synthetic data with a range of source models and atmospheric conditions. The probability, $P$, that the image contains a deformation signal according to the CNN is plotted against the maximum displacement for a range of simulations. The range of deformation signals, $D$, is created using a Monte Carlo approach and a point (Mogi) model. The deformation is combined with stratified ($S$) and turbulent ($T$) atmospheric components according to the equation $X=D+\alpha S+ \beta T$. $\alpha$ and $\beta$ are set equally to 0, 0.5, 1 in the first, second and third rows, respectively and more combinations of $\alpha$ and $\beta$ are shown in the supplementary material (Figure S1). The last row shows the detection thresholds when $P$=0.5 for different values of $\mu$ and $\tau$.}
    \label{fig:result_syn_mul_shift_plot}
\end{figure*}

 Figure \ref{fig:result_syn_mul_shift_plot} shows the output probability, $P$, against maximum displacement for each interferogram in the synthetic dataset. The transition between $P\sim0$ (undetectable) and $P\sim1$ (detectable) occurs over a narrow range and corresponds to the minimum value of displacement that the CNN can identify. We fit a sigmoidal curve, defined as $f(x)$=$(1+e^{-a(x-b)})^{-1}$, to estimate the detection threshold for each of the different levels of stratified noise ($\alpha$) and turbulent noise ($\beta$). As expected, stronger stratified atmosphere (increasing $\alpha$) and stronger turbulent atmosphere (increasing $\beta$) increases the detection threshold from 3.9 cm for $\alpha=\beta=0$ to 6.3 cm for $\alpha=\beta=1$. Conversely, increasing the wrap gain from $\mu=1$ to $\mu=2$ reduces the detection threshold by 25-30\%, e.g. for $\alpha=\beta=0.5$ and 1, giving final values of 3.8 cm and 5.2 cm respectively. 
However, although the higher wrap gains produce a lower detection threshold overall, there is significant scatter in the probabilities, suggesting that such a high wrap gain would produce a large number of false positive results.
Shifting the wrap boundaries produces a small reduction in detection threshold for noise-free data; but no change in the overall detection threshold when noise is included in the simulation as shown in Figure \ref{fig:result_syn_mul_shift_plot} bottom-row. 
 
The estimated detection thresholds can be applied either to a single interferogram or to a cumulative time-series, where they correspond to the deformation rate multiplied by the time-series duration. For example, the detection threshold of 5.2 cm for a volcanic environment corresponds to a steady rate of 1.0 cm/yr for the 4 years of Sentinel-1 data available at the time of writing (2015-2019). This is a significant improvement to the minimum rate of just 1.80~m/y reported in \citeA{Anantrasirichai:Application:2018} for applying CNNs to individual 12 day wrapped interferograms.

\subsection{Application to real examples} 

\begin{figure*}[!ht]
	\centering
      		\includegraphics[width=\textwidth]{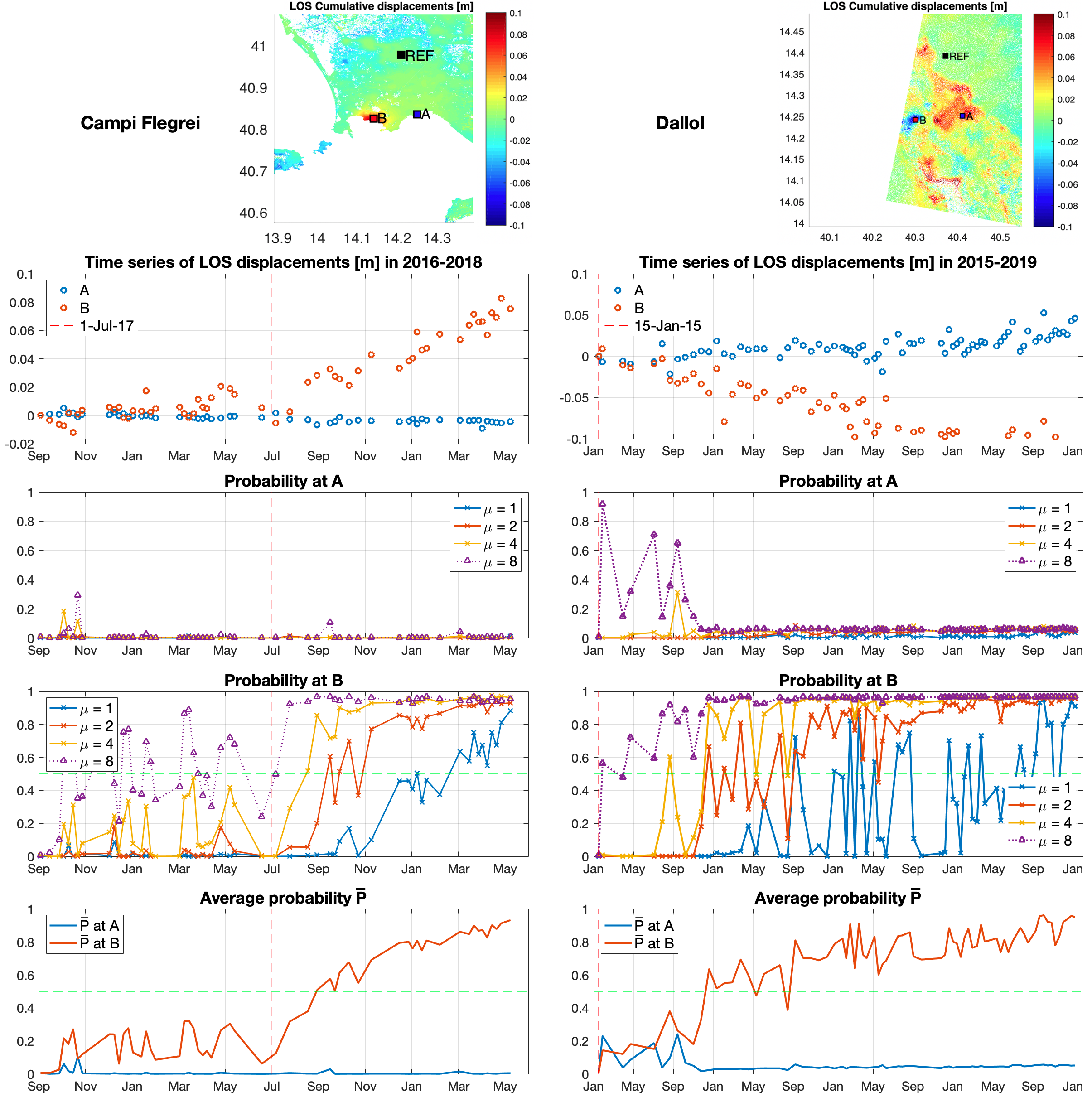}  
					\caption{Examples using Sentinel-1 time-series from Campi Flegrei (left column) and Dallol (right column). Second-row figures show the cumulative displacements and two studied areas, A and B. The ground at area A is considered stable, whereas B is located at the volcanic centre is shows significant deformation. The CNN-output probability for points A and B are shown in the third and fourth rows for wrap gains $\mu$=1,2,4,8. The fifth row contains the average probability from the selected wrap gains as defined by Equation \ref{equ:finalprob}.}
    \label{fig:result_real_plot}
\end{figure*}

For both the Campi Flegrei and Dallol datasets (Figure \ref{fig:result_real_plot}), the CNN outputs a probability map for each step of the time-series and we plot the probability values at the stable point A (Figure \ref{fig:result_real_plot} third row) and the volcanic centre, B (Figure \ref{fig:result_real_plot} forth row). For each example, we test four wrap gains, i.e. $\mu$=1, 2, 4 and 8. 

For Campi Flegrei, the probability at point A is $<$0.05 for almost all timesteps, whereas at point B, there is an increase in $P$ around the start of the deformation. 
Using the original wrap gain ($\mu$=1), we detect deformation in Feb 2018, 7 months after the episode began. At this point, the cumulative displacement is $>5$~cm, consistent with the detection threshold in our synthetic tests. 
In the times series with $\mu$ equal to 2, 4 and 8, we identify deformation at $\sim$3~cm (2.5 months), $\sim$2~cm (1.5 months) and $<$1~cm ($-$9 months) respectively. However, the higher wrap gains ($\mu>4$) incorrectly exceed the detection threshold of $P=0.5$ before the deformation began, with a total of 15 false positives for $\mu=8$.

At Dallol, the deformation time series is noisier and the rate of deformation (3.5~cm/yr) is lower than the previous example. At $\mu=1$, the probability first exceeds $P=0.5$ after 21 months, but continues to drop below the threshold throughout the entire 4 year time series (Figure \ref{fig:result_real_plot} right). The cumulative deformation at this point is 10~cm, significantly greater the synthetic detection threshold. Visual inspection shows that the low probabilities occur when the surrounding area is incoherent, a situation that was not included in the synthetic tests. In these cases, doubling the wrap gain to $\mu=2$ increases the probability above the threshold ($P=0.5$). The highest wrap gain $\mu$=8 identifies deformation at point B in only the second time step when the displacement is $<1$ cm, but it falsely identifies deformation at point A for the first 9 months. Examples of wrapped interferograms at Campi Flegrei and Dallol when $\mu$=1 and 2 are shown in Figure S2, and the interferograms causing false positives at Dallol are shown in Figure S3.

Increasing the wrap gain ($\mu$) of the interferograms reduces the detection threshold for the CNN, thus allowing us to detect slow, sustained deformation earlier in time series. However, increasing the wrap gain too much causes the CNN to misidentify features caused by atmospheric artefacts. Consequently, decisions based on only large values of $\mu$ would have many false positives, while those using only small values of $\mu$ might not be able to detect slow deformation. We therefore compute the final probability $\overline{P}$ by combining the probability at a range of wrap gains (Equation \ref{equ:finalprob}), using $N$=4.

\begin{linenomath*}
\begin{equation}
\label{equ:finalprob}
\overline{P} = \frac{1}{N} \sum_{i=0}^{N-1} P_{\mu=2^i}.
\end{equation}
\end{linenomath*}

 The value of $\overline{P}$ shows a similar trend to the magnitude of displacement -- low  $\overline{P}$ at small displacement and large  $\overline{P}$ at large displacement (Figure \ref{fig:result_real_plot} last row).  
 The probability first exceeds the threshold $\overline{P}=0.5$ on 27 August 2017 for Campi Flegrei and 2 December 2015 for Dallol (Figure \ref{fig:result_real_plot}), corresponding to 2 months after the onset of deformation and 11 months after the start of the timeseries respectively. 

\subsection{Discussion}

\begin{figure*}[!ht]
	\centering
      		\includegraphics[width=0.8\textwidth]{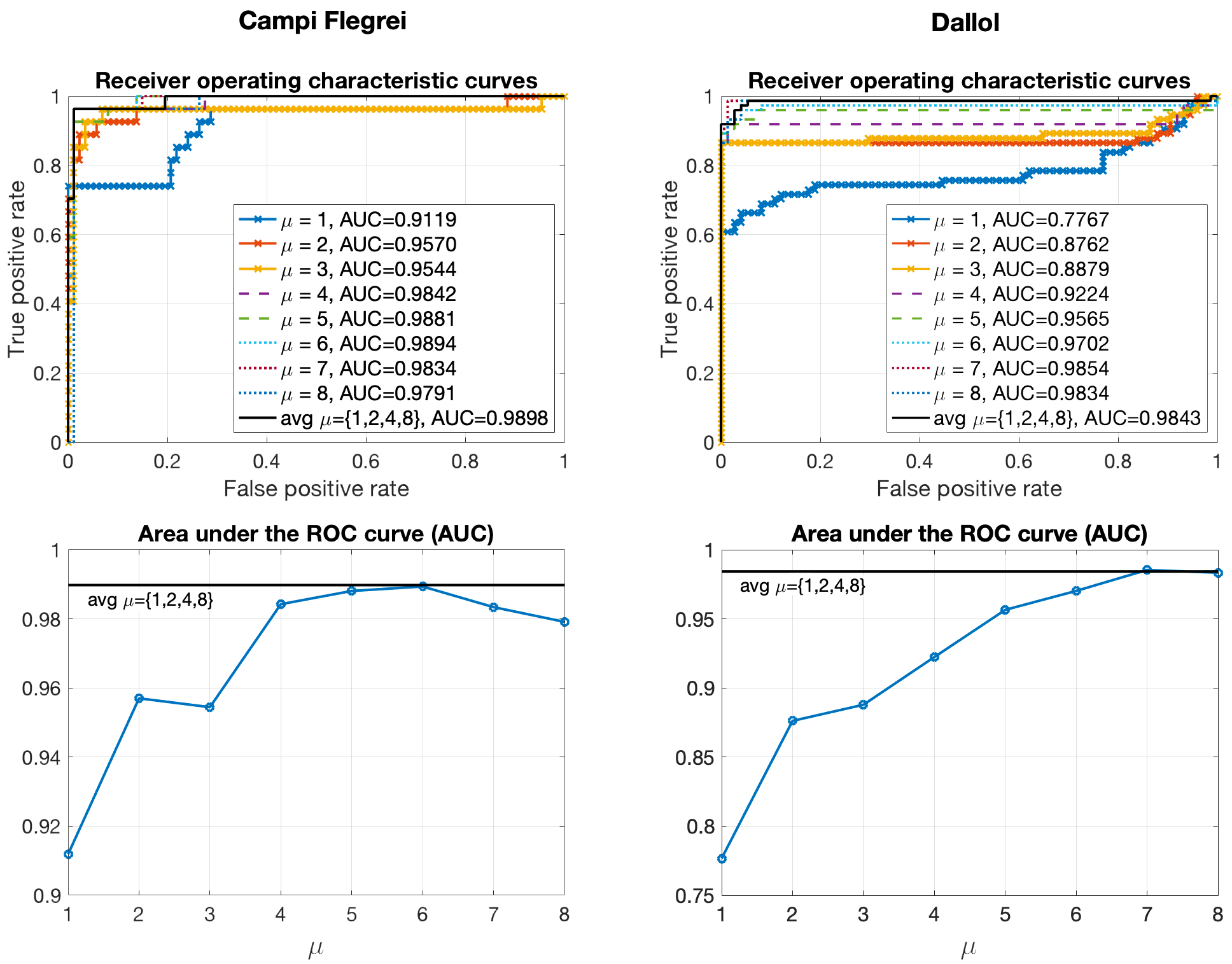}  
					\caption{Performance analysis of the CNN approach.  The top row shows the ROC curves for the points A and B, whilst the bottom row demonstrates the area under the ROC curves (AUC) for wrap gains $\mu$=1,2,4,8.  The higher AUC indicates better performance (fewer false positives and false negatives).  (Left column) Campi Flegrei and (right column) Dallol. The start date of deformation is taken to be 1 July 2017 for Campi Flegrei and 15 January 2015 for Dallol, which corresponds to the start of the time series.}
    \label{fig:result_performance}
\end{figure*}

We evaluated the results from the examples at Campi Flegrei and Dallol using a receiver operating characteristic (ROC) curve  (Figure \ref{fig:result_performance}), where true positive (TPR) and false positive rates (FPR) were calculated by varying the probability thresholds for identify deformation and non-deformation. The TPR is the fraction of predicted deformation that are retrieved over  the total number of actual deformation, whilst the FPR is the number of non-deformation wrongly identified as deformation divided by the total number of actual non-deformation. We computed the area under the ROC curve (AUC): good classifiers will give high AUC values as they detect the positive signals correctly and few true negatives are falsely identified. For Campi Flegrei, the highest AUC (0.989) is produced by $\mu$=6, which indicates good separation between classes. The AUC decreases for higher wrap gains as the number of false positives increases. For Dallol, the ROC curves are computed assuming the deformation starts on the first date of the time series (15th January 2015) and the highest AUC results (0.985) are for wrap gains of $\mu$=7. The AUC under the curve for the combined probability $\overline{P}$ is 0.989 and 0.984 for  Campi Flegrei and Dallol datasets, respectively. This demonstrates that CNNs can be applied to wrapped InSAR data to detect deformation at rates as low as 3.5~cm/yr.

Here, we have tested the ability of CNNs to detect slow, steady deformation in volcanic environments, but further testing is still required to assess whether this technique could also be used to detect anthropogenic sources of deformation, such as subsidence associated with resource extraction or infrastructure instability. The characteristics of the deformation are quite different, but so are the noise characteristics. In particular, the deformation has a smaller spatial extent and the topographic gradients smaller, which means that it is possible to reduce the atmospheric components by filtering \cite{Hooper:new:2004,Ferretti:new:2011}. Thus, although the rates of deformation are lower, we might expect similar signal-to-noise ratios.

Our approach is based on the ability of CNNs to distinguish between different sources using the pattern of fringes \cite{Anantrasirichai:Application:2018,Anantrasirichai:deep:2019}. However, to create cumulative time series, the interferograms need to be unwrapped first, and we artificially rewrap the data. Alternatively CNN-based methods applied directly to unwrapped data may also have the potential to detect slow, steady deformation and a direct comparison of the performance of the two approaches using robust performance metrics on a suitable range of case studies would be interesting.

\section{Conclusions}
In this paper, we study the feasibility of using convolutional neural networks (CNNs) to detect slow surface deformation in InSAR images, by rewrapping cumulative time series. We first tested the technique on synthetic interferograms based on volcanic environments, and found that when the number of fringes is doubled, the detection threshold is lowered by 25-30\%. The detection threshold of 5.2 cm corresponds to rates of 1.3 cm/yr for the 4 year Sentinel-1 time series currently available. Subsequently we test using the cumulative times series from two volcanoes: Campi Flegrei and Dallol and demonstrate the ability to detect rates as low as 3.5 cm/y, with an improvement of the AUC up to 15\%  when over-wrapping is applied. However, increasing the wrap gain too high increases the number of false negatives, so we propose a method to combine the results from multiple wrapping gains. This shows potential for future use, though further adaptations may be required to apply the technique to a wider range of environments.


\acknowledgments
This work was supported by the  NERC BGS Centre for Observation and Modelling of Earthquakes, Volcanoes and Tectonics (COMET), the NERC large grant Looking into the Continents from Space (NE/K010913/1), NERC innovation - Making Satellite Volcano Deformation Analysis Accessible (NE/S013970/1), and the EPSRC Platform Grant - Vision for the Future (EP/M000885/1). The InSAR datasets are available at \url{http://comet.nerc.ac.uk/COMET-LiCS-portal/}.

%

\begin{thebibliography}{}

\bibitem [\protect \citeauthoryear {%
Anantrasirichai%
, Biggs%
, Albino%
\BCBL {}\ \BBA {} Bull%
}{%
Anantrasirichai%
\ \protect \BOthers {.}}{%
{\protect \APACyear {2019}}%
}]{%
Anantrasirichai:deep:2019}
\APACinsertmetastar {%
Anantrasirichai:deep:2019}%
\begin{APACrefauthors}%
Anantrasirichai, N.%
, Biggs, J.%
, Albino, F.%
\BCBL {}\ \BBA {} Bull, D.%
\end{APACrefauthors}%
\unskip\
\newblock
\APACrefYearMonthDay{2019}{}{}.
\newblock
{\BBOQ}\APACrefatitle {A deep learning approach to detecting volcano
  deformation from satellite imagery using synthetic datasets} {A deep learning
  approach to detecting volcano deformation from satellite imagery using
  synthetic datasets}.{\BBCQ}
\newblock
\APACjournalVolNumPages{Remote Sensing of Environment}{230}{}{111179}.
\newblock
\begin{APACrefDOI} \doi{https://doi.org/10.1016/j.rse.2019.04.032}
  \end{APACrefDOI}
\PrintBackRefs{\CurrentBib}

\bibitem [\protect \citeauthoryear {%
Anantrasirichai%
, Biggs%
, Albino%
, Hill%
\BCBL {}\ \BBA {} Bull%
}{%
Anantrasirichai%
\ \protect \BOthers {.}}{%
{\protect \APACyear {2018}}%
}]{%
Anantrasirichai:Application:2018}
\APACinsertmetastar {%
Anantrasirichai:Application:2018}%
\begin{APACrefauthors}%
Anantrasirichai, N.%
, Biggs, J.%
, Albino, F.%
, Hill, P.%
\BCBL {}\ \BBA {} Bull, D.%
\end{APACrefauthors}%
\unskip\
\newblock
\APACrefYearMonthDay{2018}{August}{}.
\newblock
{\BBOQ}\APACrefatitle {Application of Machine Learning to Classification of
  Volcanic Deformation in Routinely-Generated In{SAR} data} {Application of
  machine learning to classification of volcanic deformation in
  routinely-generated in{SAR} data}.{\BBCQ}
\newblock
\APACjournalVolNumPages{Journal of Geophysical Research: Solid
  Earth}{123}{8}{6592-6606}.
\newblock
\begin{APACrefDOI} \doi{10.1029/2018JB015911} \end{APACrefDOI}
\PrintBackRefs{\CurrentBib}

\bibitem [\protect \citeauthoryear {%
Beauducel%
, Briole%
\BCBL {}\ \BBA {} Froger%
}{%
Beauducel%
\ \protect \BOthers {.}}{%
{\protect \APACyear {2000}}%
}]{%
Beauducel:volcano:2000}
\APACinsertmetastar {%
Beauducel:volcano:2000}%
\begin{APACrefauthors}%
Beauducel, F.%
, Briole, P.%
\BCBL {}\ \BBA {} Froger, J\BHBI L.%
\end{APACrefauthors}%
\unskip\
\newblock
\APACrefYearMonthDay{2000}{}{}.
\newblock
{\BBOQ}\APACrefatitle {Volcano-wide fringes in ERS synthetic aperture radar
  interferograms of Etna (1992--1998): Deformation or tropospheric effect?}
  {Volcano-wide fringes in ers synthetic aperture radar interferograms of etna
  (1992--1998): Deformation or tropospheric effect?}{\BBCQ}
\newblock
\APACjournalVolNumPages{Journal of Geophysical Research: Solid
  Earth}{105}{B7}{16391--16402}.
\PrintBackRefs{\CurrentBib}

\bibitem [\protect \citeauthoryear {%
Biggs%
\ \BBA {} Pritchard%
}{%
Biggs%
\ \BBA {} Pritchard%
}{%
{\protect \APACyear {2017}}%
}]{%
biggs2017global}
\APACinsertmetastar {%
biggs2017global}%
\begin{APACrefauthors}%
Biggs, J.%
\BCBT {}\ \BBA {} Pritchard, M\BPBI E.%
\end{APACrefauthors}%
\unskip\
\newblock
\APACrefYearMonthDay{2017}{}{}.
\newblock
{\BBOQ}\APACrefatitle {Global volcano monitoring: what does it mean when
  volcanoes deform?} {Global volcano monitoring: what does it mean when
  volcanoes deform?}{\BBCQ}
\newblock
\APACjournalVolNumPages{Elements}{13}{1}{17--22}.
\newblock
\begin{APACrefDOI} \doi{10.2113/gselements.13.1.17} \end{APACrefDOI}
\PrintBackRefs{\CurrentBib}

\bibitem [\protect \citeauthoryear {%
Biggs%
, Wright%
, Lu%
\BCBL {}\ \BBA {} Parsons%
}{%
Biggs%
\ \protect \BOthers {.}}{%
{\protect \APACyear {2007}}%
}]{%
biggs2007multi}
\APACinsertmetastar {%
biggs2007multi}%
\begin{APACrefauthors}%
Biggs, J.%
, Wright, T.%
, Lu, Z.%
\BCBL {}\ \BBA {} Parsons, B.%
\end{APACrefauthors}%
\unskip\
\newblock
\APACrefYearMonthDay{2007}{}{}.
\newblock
{\BBOQ}\APACrefatitle {Multi-interferogram method for measuring interseismic
  deformation: Denali {F}ault, {A}laska} {Multi-interferogram method for
  measuring interseismic deformation: Denali {F}ault, {A}laska}.{\BBCQ}
\newblock
\APACjournalVolNumPages{Geophysical Journal International}{170}{3}{1165--1179}.
\newblock
\begin{APACrefDOI} \doi{10.1111/j.1365-246X.2007.03415.x} \end{APACrefDOI}
\PrintBackRefs{\CurrentBib}

\bibitem [\protect \citeauthoryear {%
{Chen}%
\ \BBA {} {Zebker}%
}{%
{Chen}%
\ \BBA {} {Zebker}%
}{%
{\protect \APACyear {2002}}%
}]{%
Chen:phase:2002}
\APACinsertmetastar {%
Chen:phase:2002}%
\begin{APACrefauthors}%
{Chen}, C\BPBI W.%
\BCBT {}\ \BBA {} {Zebker}, H\BPBI A.%
\end{APACrefauthors}%
\unskip\
\newblock
\APACrefYearMonthDay{2002}{Aug}{}.
\newblock
{\BBOQ}\APACrefatitle {Phase unwrapping for large SAR interferograms:
  statistical segmentation and generalized network models} {Phase unwrapping
  for large sar interferograms: statistical segmentation and generalized
  network models}.{\BBCQ}
\newblock
\APACjournalVolNumPages{IEEE Transactions on Geoscience and Remote
  Sensing}{40}{8}{1709-1719}.
\newblock
\begin{APACrefDOI} \doi{10.1109/TGRS.2002.802453} \end{APACrefDOI}
\PrintBackRefs{\CurrentBib}

\bibitem [\protect \citeauthoryear {%
Del~Gaudio%
, Aquino%
, Ricciardi%
, Ricco%
\BCBL {}\ \BBA {} Scandone%
}{%
Del~Gaudio%
\ \protect \BOthers {.}}{%
{\protect \APACyear {2010}}%
}]{%
delgaudio2010unrest}
\APACinsertmetastar {%
delgaudio2010unrest}%
\begin{APACrefauthors}%
Del~Gaudio, C.%
, Aquino, I.%
, Ricciardi, G.%
, Ricco, C.%
\BCBL {}\ \BBA {} Scandone, R.%
\end{APACrefauthors}%
\unskip\
\newblock
\APACrefYearMonthDay{2010}{}{}.
\newblock
{\BBOQ}\APACrefatitle {Unrest episodes at Campi Flegrei: A reconstruction of
  vertical ground movements during 1905--2009} {Unrest episodes at campi
  flegrei: A reconstruction of vertical ground movements during
  1905--2009}.{\BBCQ}
\newblock
\APACjournalVolNumPages{Journal of Volcanology and Geothermal
  Research}{195}{1}{48--56}.
\PrintBackRefs{\CurrentBib}

\bibitem [\protect \citeauthoryear {%
Ebmeier%
, Biggs%
, Mather%
\BCBL {}\ \BBA {} Amelung%
}{%
Ebmeier%
\ \protect \BOthers {.}}{%
{\protect \APACyear {2013}}%
}]{%
Ebmeier:Applicability:2013}
\APACinsertmetastar {%
Ebmeier:Applicability:2013}%
\begin{APACrefauthors}%
Ebmeier, S\BPBI K.%
, Biggs, J.%
, Mather, T\BPBI A.%
\BCBL {}\ \BBA {} Amelung, F.%
\end{APACrefauthors}%
\unskip\
\newblock
\APACrefYearMonthDay{2013}{}{}.
\newblock
{\BBOQ}\APACrefatitle {Applicability of {I}n{SAR} to tropical volcanoes:
  insights from {C}entral {A}merica} {Applicability of {I}n{SAR} to tropical
  volcanoes: insights from {C}entral {A}merica}.{\BBCQ}
\newblock
\APACjournalVolNumPages{Geological Society, London, Special
  Publications}{380}{}{15-37}.
\newblock
\begin{APACrefDOI} \doi{10.1144/SP380.2} \end{APACrefDOI}
\PrintBackRefs{\CurrentBib}

\bibitem [\protect \citeauthoryear {%
{Ferretti}%
\ \protect \BOthers {.}}{%
{Ferretti}%
\ \protect \BOthers {.}}{%
{\protect \APACyear {2011}}%
}]{%
Ferretti:new:2011}
\APACinsertmetastar {%
Ferretti:new:2011}%
\begin{APACrefauthors}%
{Ferretti}, A.%
, {Fumagalli}, A.%
, {Novali}, F.%
, {Prati}, C.%
, {Rocca}, F.%
\BCBL {}\ \BBA {} {Rucci}, A.%
\end{APACrefauthors}%
\unskip\
\newblock
\APACrefYearMonthDay{2011}{Sep.}{}.
\newblock
{\BBOQ}\APACrefatitle {A New Algorithm for Processing Interferometric
  Data-Stacks: SqueeSAR} {A new algorithm for processing interferometric
  data-stacks: Squeesar}.{\BBCQ}
\newblock
\APACjournalVolNumPages{IEEE Transactions on Geoscience and Remote
  Sensing}{49}{9}{3460-3470}.
\newblock
\begin{APACrefDOI} \doi{10.1109/TGRS.2011.2124465} \end{APACrefDOI}
\PrintBackRefs{\CurrentBib}

\bibitem [\protect \citeauthoryear {%
{Global Volcanism Program}%
}{%
{Global Volcanism Program}%
}{%
{\protect \APACyear {2013}}%
}]{%
global2013volcanoes}
\APACinsertmetastar {%
global2013volcanoes}%
\begin{APACrefauthors}%
{Global Volcanism Program}.%
\end{APACrefauthors}%
\unskip\
\newblock
\APACrefYearMonthDay{2013}{}{}.
\newblock
{\BBOQ}\APACrefatitle {Volcanoes of the World} {Volcanoes of the world}.{\BBCQ}
\newblock
\APACjournalVolNumPages{Smithsonian Institution}{}{}{}.
\newblock
\begin{APACrefDOI} \doi{https://doi.org/10.5479/si.GVP.VOTW4-2013}
  \end{APACrefDOI}
\PrintBackRefs{\CurrentBib}

\bibitem [\protect \citeauthoryear {%
{Global Volcanism Program}%
}{%
{Global Volcanism Program}%
}{%
{\protect \APACyear {2019}}%
}]{%
GVP2019}
\APACinsertmetastar {%
GVP2019}%
\begin{APACrefauthors}%
{Global Volcanism Program}.%
\end{APACrefauthors}%
\unskip\
\newblock
\APACrefYearMonthDay{2019}{}{}.
\newblock
{\BBOQ}\APACrefatitle {Piton de la {F}ournaise ({F}rance) Eruptive episodes in
  {F}ebruary-{M}arch and {J}une 2019; multiple fissures and lava flows} {Piton
  de la {F}ournaise ({F}rance) eruptive episodes in {F}ebruary-{M}arch and
  {J}une 2019; multiple fissures and lava flows}.{\BBCQ}
\newblock
\APACjournalVolNumPages{}{44}{7}{}.
\PrintBackRefs{\CurrentBib}

\bibitem [\protect \citeauthoryear {%
Henderson%
\ \BBA {} Pritchard%
}{%
Henderson%
\ \BBA {} Pritchard%
}{%
{\protect \APACyear {2017}}%
}]{%
Henderson:time:2017}
\APACinsertmetastar {%
Henderson:time:2017}%
\begin{APACrefauthors}%
Henderson, S\BPBI T.%
\BCBT {}\ \BBA {} Pritchard, M\BPBI E.%
\end{APACrefauthors}%
\unskip\
\newblock
\APACrefYearMonthDay{2017}{}{}.
\newblock
{\BBOQ}\APACrefatitle {Time-dependent deformation of Uturuncu volcano, Bolivia,
  constrained by GPS and InSAR measurements and implications for source models}
  {Time-dependent deformation of uturuncu volcano, bolivia, constrained by gps
  and insar measurements and implications for source models}.{\BBCQ}
\newblock
\APACjournalVolNumPages{Geosphere}{13}{6}{1834--1854}.
\PrintBackRefs{\CurrentBib}

\bibitem [\protect \citeauthoryear {%
Hooper%
, Zebker%
, Segall%
\BCBL {}\ \BBA {} Kampes%
}{%
Hooper%
\ \protect \BOthers {.}}{%
{\protect \APACyear {2004}}%
}]{%
Hooper:new:2004}
\APACinsertmetastar {%
Hooper:new:2004}%
\begin{APACrefauthors}%
Hooper, A.%
, Zebker, H.%
, Segall, P.%
\BCBL {}\ \BBA {} Kampes, B.%
\end{APACrefauthors}%
\unskip\
\newblock
\APACrefYearMonthDay{2004}{}{}.
\newblock
{\BBOQ}\APACrefatitle {A new method for measuring deformation on volcanoes and
  other natural terrains using InSAR persistent scatterers} {A new method for
  measuring deformation on volcanoes and other natural terrains using insar
  persistent scatterers}.{\BBCQ}
\newblock
\APACjournalVolNumPages{Geophysical Research Letters}{31}{23}{}.
\newblock
\begin{APACrefURL}
  \url{https://agupubs.onlinelibrary.wiley.com/doi/abs/10.1029/2004GL021737}
  \end{APACrefURL}
\newblock
\begin{APACrefDOI} \doi{10.1029/2004GL021737} \end{APACrefDOI}
\PrintBackRefs{\CurrentBib}

\bibitem [\protect \citeauthoryear {%
Lloyd%
\ \protect \BOthers {.}}{%
Lloyd%
\ \protect \BOthers {.}}{%
{\protect \APACyear {2018}}%
}]{%
Lloyd:Evidence:2018}
\APACinsertmetastar {%
Lloyd:Evidence:2018}%
\begin{APACrefauthors}%
Lloyd, R.%
, Biggs, J.%
, Wilks, M.%
, Nowacki, A.%
, Kendall, J\BPBI M.%
, Ayele, A.%
\BDBL {}Eysteinsson, H.%
\end{APACrefauthors}%
\unskip\
\newblock
\APACrefYearMonthDay{2018}{}{}.
\newblock
{\BBOQ}\APACrefatitle {Evidence for cross rift structural controls on
  deformation and seismicity at a continental rift caldera} {Evidence for cross
  rift structural controls on deformation and seismicity at a continental rift
  caldera}.{\BBCQ}
\newblock
\APACjournalVolNumPages{Earth and Planetary Science Letters}{487}{}{190-200}.
\newblock
\begin{APACrefDOI} \doi{https://doi.org/10.1016/j.epsl.2018.01.037}
  \end{APACrefDOI}
\PrintBackRefs{\CurrentBib}

\bibitem [\protect \citeauthoryear {%
McCay%
, Valyrakis%
\BCBL {}\ \BBA {} Younger%
}{%
McCay%
\ \protect \BOthers {.}}{%
{\protect \APACyear {2018}}%
}]{%
McCay:meta:2018}
\APACinsertmetastar {%
McCay:meta:2018}%
\begin{APACrefauthors}%
McCay, A\BPBI T.%
, Valyrakis, M.%
\BCBL {}\ \BBA {} Younger, P\BPBI L.%
\end{APACrefauthors}%
\unskip\
\newblock
\APACrefYearMonthDay{2018}{}{}.
\newblock
{\BBOQ}\APACrefatitle {A meta-analysis of coal mining induced subsidence data
  and implications for their use in the carbon industry} {A meta-analysis of
  coal mining induced subsidence data and implications for their use in the
  carbon industry}.{\BBCQ}
\newblock
\APACjournalVolNumPages{International Journal of Coal Geology}{192}{}{91-101}.
\newblock
\begin{APACrefDOI} \doi{https://doi.org/10.1016/j.coal.2018.03.013}
  \end{APACrefDOI}
\PrintBackRefs{\CurrentBib}

\bibitem [\protect \citeauthoryear {%
Mogi%
}{%
Mogi%
}{%
{\protect \APACyear {1958}}%
}]{%
Mogi:Relation:1958}
\APACinsertmetastar {%
Mogi:Relation:1958}%
\begin{APACrefauthors}%
Mogi, K.%
\end{APACrefauthors}%
\unskip\
\newblock
\APACrefYearMonthDay{1958}{}{}.
\newblock
{\BBOQ}\APACrefatitle {Relation between the eruptions of various volcanoes and
  deformations of the ground surfaces around them} {Relation between the
  eruptions of various volcanoes and deformations of the ground surfaces around
  them}.{\BBCQ}
\newblock
\APACjournalVolNumPages{Bull. Earthquake Res. Inst. Univ. Tokyo}{36}{}{99-134}.
\PrintBackRefs{\CurrentBib}

\bibitem [\protect \citeauthoryear {%
Montgomery-Brown%
\ \protect \BOthers {.}}{%
Montgomery-Brown%
\ \protect \BOthers {.}}{%
{\protect \APACyear {2015}}%
}]{%
Montgomery:renewed:2015}
\APACinsertmetastar {%
Montgomery:renewed:2015}%
\begin{APACrefauthors}%
Montgomery-Brown, E.%
, Wicks, C.%
, Cervelli, P\BPBI F.%
, Langbein, J\BPBI O.%
, Svarc, J\BPBI L.%
, Shelly, D\BPBI R.%
\BDBL {}Lisowski, M.%
\end{APACrefauthors}%
\unskip\
\newblock
\APACrefYearMonthDay{2015}{}{}.
\newblock
{\BBOQ}\APACrefatitle {Renewed inflation of Long Valley Caldera, California
  (2011 to 2014)} {Renewed inflation of long valley caldera, california (2011
  to 2014)}.{\BBCQ}
\newblock
\APACjournalVolNumPages{Geophysical Research Letters}{42}{13}{5250--5257}.
\PrintBackRefs{\CurrentBib}

\bibitem [\protect \citeauthoryear {%
Nobile%
\ \protect \BOthers {.}}{%
Nobile%
\ \protect \BOthers {.}}{%
{\protect \APACyear {2012}}%
}]{%
nobile2012dike}
\APACinsertmetastar {%
nobile2012dike}%
\begin{APACrefauthors}%
Nobile, A.%
, Pagli, C.%
, Keir, D.%
, Wright, T\BPBI J.%
, Ayele, A.%
, Ruch, J.%
\BCBL {}\ \BBA {} Acocella, V.%
\end{APACrefauthors}%
\unskip\
\newblock
\APACrefYearMonthDay{2012}{}{}.
\newblock
{\BBOQ}\APACrefatitle {Dike-fault interaction during the 2004 Dallol intrusion
  at the northern edge of the Erta Ale Ridge (Afar, Ethiopia)} {Dike-fault
  interaction during the 2004 dallol intrusion at the northern edge of the erta
  ale ridge (afar, ethiopia)}.{\BBCQ}
\newblock
\APACjournalVolNumPages{Geophysical Research Letters}{39}{19}{}.
\PrintBackRefs{\CurrentBib}

\bibitem [\protect \citeauthoryear {%
Remy%
\ \protect \BOthers {.}}{%
Remy%
\ \protect \BOthers {.}}{%
{\protect \APACyear {2015}}%
}]{%
Remy:revised:2015}
\APACinsertmetastar {%
Remy:revised:2015}%
\begin{APACrefauthors}%
Remy, D.%
, Chen, Y.%
, Froger, J\BHBI L.%
, Bonvalot, S.%
, Cordoba, L.%
\BCBL {}\ \BBA {} Fustos, J.%
\end{APACrefauthors}%
\unskip\
\newblock
\APACrefYearMonthDay{2015}{}{}.
\newblock
{\BBOQ}\APACrefatitle {Revised interpretation of recent InSAR signals observed
  at Llaima volcano (Chile)} {Revised interpretation of recent insar signals
  observed at llaima volcano (chile)}.{\BBCQ}
\newblock
\APACjournalVolNumPages{Geophysical Research Letters}{42}{10}{3870--3879}.
\PrintBackRefs{\CurrentBib}

\bibitem [\protect \citeauthoryear {%
Remy%
\ \protect \BOthers {.}}{%
Remy%
\ \protect \BOthers {.}}{%
{\protect \APACyear {2014}}%
}]{%
Remy:persistent:2014}
\APACinsertmetastar {%
Remy:persistent:2014}%
\begin{APACrefauthors}%
Remy, D.%
, Froger, J\BHBI L.%
, Perfettini, H.%
, Bonvalot, S.%
, Gabalda, G.%
, Albino, F.%
\BDBL {}Saint~Blanquat, M\BPBI D.%
\end{APACrefauthors}%
\unskip\
\newblock
\APACrefYearMonthDay{2014}{}{}.
\newblock
{\BBOQ}\APACrefatitle {Persistent uplift of the L azufre volcanic complex (C
  entral A ndes): New insights from PCAIM inversion of InSAR time series and
  GPS data} {Persistent uplift of the l azufre volcanic complex (c entral a
  ndes): New insights from pcaim inversion of insar time series and gps
  data}.{\BBCQ}
\newblock
\APACjournalVolNumPages{Geochemistry, Geophysics,
  Geosystems}{15}{9}{3591--3611}.
\PrintBackRefs{\CurrentBib}

\bibitem [\protect \citeauthoryear {%
Schmidt%
\ \BBA {} B{\"u}rgmann%
}{%
Schmidt%
\ \BBA {} B{\"u}rgmann%
}{%
{\protect \APACyear {2003}}%
}]{%
schmidt2003time}
\APACinsertmetastar {%
schmidt2003time}%
\begin{APACrefauthors}%
Schmidt, D\BPBI A.%
\BCBT {}\ \BBA {} B{\"u}rgmann, R.%
\end{APACrefauthors}%
\unskip\
\newblock
\APACrefYearMonthDay{2003}{}{}.
\newblock
{\BBOQ}\APACrefatitle {Time-dependent land uplift and subsidence in the Santa
  Clara valley, California, from a large interferometric synthetic aperture
  radar data set} {Time-dependent land uplift and subsidence in the santa clara
  valley, california, from a large interferometric synthetic aperture radar
  data set}.{\BBCQ}
\newblock
\APACjournalVolNumPages{Journal of Geophysical Research: Solid
  Earth}{108}{B9}{}.
\PrintBackRefs{\CurrentBib}

\bibitem [\protect \citeauthoryear {%
Trasatti%
\ \protect \BOthers {.}}{%
Trasatti%
\ \protect \BOthers {.}}{%
{\protect \APACyear {2008}}%
}]{%
Trasatti:uplift:2008}
\APACinsertmetastar {%
Trasatti:uplift:2008}%
\begin{APACrefauthors}%
Trasatti, E.%
, Casu, F.%
, Giunchi, C.%
, Pepe, S.%
, Solaro, G.%
, Tagliaventi, S.%
\BDBL {}others%
\end{APACrefauthors}%
\unskip\
\newblock
\APACrefYearMonthDay{2008}{}{}.
\newblock
{\BBOQ}\APACrefatitle {The 2004--2006 uplift episode at Campi Flegrei caldera
  (Italy): Constraints from SBAS-DInSAR ENVISAT data and Bayesian source
  inference} {The 2004--2006 uplift episode at campi flegrei caldera (italy):
  Constraints from sbas-dinsar envisat data and bayesian source
  inference}.{\BBCQ}
\newblock
\APACjournalVolNumPages{Geophysical Research Letters}{35}{7}{}.
\PrintBackRefs{\CurrentBib}

\bibitem [\protect \citeauthoryear {%
Troiano%
, Di~Giuseppe%
, Petrillo%
, Troise%
\BCBL {}\ \BBA {} De~Natale%
}{%
Troiano%
\ \protect \BOthers {.}}{%
{\protect \APACyear {2011}}%
}]{%
troiano2011ground}
\APACinsertmetastar {%
troiano2011ground}%
\begin{APACrefauthors}%
Troiano, A.%
, Di~Giuseppe, M\BPBI G.%
, Petrillo, Z.%
, Troise, C.%
\BCBL {}\ \BBA {} De~Natale, G.%
\end{APACrefauthors}%
\unskip\
\newblock
\APACrefYearMonthDay{2011}{}{}.
\newblock
{\BBOQ}\APACrefatitle {Ground deformation at calderas driven by fluid
  injection: modelling unrest episodes at Campi Flegrei (Italy)} {Ground
  deformation at calderas driven by fluid injection: modelling unrest episodes
  at campi flegrei (italy)}.{\BBCQ}
\newblock
\APACjournalVolNumPages{Geophysical Journal International}{187}{2}{833--847}.
\PrintBackRefs{\CurrentBib}

\bibitem [\protect \citeauthoryear {%
Usai%
}{%
Usai%
}{%
{\protect \APACyear {2003}}%
}]{%
usai2003least}
\APACinsertmetastar {%
usai2003least}%
\begin{APACrefauthors}%
Usai, S.%
\end{APACrefauthors}%
\unskip\
\newblock
\APACrefYearMonthDay{2003}{}{}.
\newblock
{\BBOQ}\APACrefatitle {A least squares database approach for SAR
  interferometric data} {A least squares database approach for sar
  interferometric data}.{\BBCQ}
\newblock
\APACjournalVolNumPages{IEEE Transactions on Geoscience and Remote
  Sensing}{41}{4}{753--760}.
\PrintBackRefs{\CurrentBib}

\bibitem [\protect \citeauthoryear {%
Valade%
\ \protect \BOthers {.}}{%
Valade%
\ \protect \BOthers {.}}{%
{\protect \APACyear {2019}}%
}]{%
Valade:towards:2019}
\APACinsertmetastar {%
Valade:towards:2019}%
\begin{APACrefauthors}%
Valade, S.%
, Ley, A.%
, Massimetti, F.%
, D'Hondt, O.%
, Laiolo, M.%
, Coppola, D.%
\BDBL {}Walter, T\BPBI R.%
\end{APACrefauthors}%
\unskip\
\newblock
\APACrefYearMonthDay{2019}{}{}.
\newblock
{\BBOQ}\APACrefatitle {Towards Global Volcano Monitoring Using Multisensor
  Sentinel Missions and Artificial Intelligence: The MOUNTS Monitoring System}
  {Towards global volcano monitoring using multisensor sentinel missions and
  artificial intelligence: The mounts monitoring system}.{\BBCQ}
\newblock
\APACjournalVolNumPages{Remote Sensing}{11}{13}{}.
\newblock
\begin{APACrefURL} \url{https://www.mdpi.com/2072-4292/11/13/1528}
  \end{APACrefURL}
\newblock
\begin{APACrefDOI} \doi{10.3390/rs11131528} \end{APACrefDOI}
\PrintBackRefs{\CurrentBib}

\bibitem [\protect \citeauthoryear {%
Whiteley%
, Chambers%
, Uhlemann%
, Wilkinson%
\BCBL {}\ \BBA {} Kendall%
}{%
Whiteley%
\ \protect \BOthers {.}}{%
{\protect \APACyear {2019}}%
}]{%
Whiteley:Geophysical:2019}
\APACinsertmetastar {%
Whiteley:Geophysical:2019}%
\begin{APACrefauthors}%
Whiteley, J\BPBI S.%
, Chambers, J\BPBI E.%
, Uhlemann, S.%
, Wilkinson, P\BPBI B.%
\BCBL {}\ \BBA {} Kendall, J\BPBI M.%
\end{APACrefauthors}%
\unskip\
\newblock
\APACrefYearMonthDay{2019}{}{}.
\newblock
{\BBOQ}\APACrefatitle {Geophysical Monitoring of Moisture-Induced Landslides: A
  Review} {Geophysical monitoring of moisture-induced landslides: A
  review}.{\BBCQ}
\newblock
\APACjournalVolNumPages{Reviews of Geophysics}{57}{1}{106-145}.
\newblock
\begin{APACrefDOI} \doi{10.1029/2018RG000603} \end{APACrefDOI}
\PrintBackRefs{\CurrentBib}

\bibitem [\protect \citeauthoryear {%
Yu%
, Li%
, Penna%
\BCBL {}\ \BBA {} Crippa%
}{%
Yu%
\ \protect \BOthers {.}}{%
{\protect \APACyear {2018}}%
}]{%
yu2018generic}
\APACinsertmetastar {%
yu2018generic}%
\begin{APACrefauthors}%
Yu, C.%
, Li, Z.%
, Penna, N\BPBI T.%
\BCBL {}\ \BBA {} Crippa, P.%
\end{APACrefauthors}%
\unskip\
\newblock
\APACrefYearMonthDay{2018}{}{}.
\newblock
{\BBOQ}\APACrefatitle {Generic atmospheric correction model for
  {Interferometric Synthetic Aperture Radar} observations} {Generic atmospheric
  correction model for {Interferometric Synthetic Aperture Radar}
  observations}.{\BBCQ}
\newblock
\APACjournalVolNumPages{Journal of Geophysical Research: Solid Earth}{}{}{}.
\PrintBackRefs{\CurrentBib}

\end{thebibliography}
%

\end{document}